\documentclass{emulateapj}

\slugcomment{Draft date: February 23, 2008}

\shorttitle{Orbital Modulation of GX 9+9}
\shortauthors{Harris et al.}

\begin{document}


\title{GX 9+9: Variability of the X-ray Orbital Modulation}


\author{Robert J. Harris\altaffilmark{1,2}, Alan
 M. Levine\altaffilmark{3}, Martin
 Durant\altaffilmark{4}, Albert K. H. Kong\altaffilmark{5},
 Phil Charles\altaffilmark{6}, and Tariq Shahbaz\altaffilmark{4}}

\altaffiltext{1}{Department of Physics and Kavli Institute for
 Astrophysics and Space Research, MIT, Cambridge, MA 02139.}

\altaffiltext{2}{present address: Center for Astrophysics, 60 Garden
 Street, Cambridge, MA 02138; rjharris@cfa.harvard.edu.}

\altaffiltext{3}{Kavli Institute for Astrophysics and Space Research,
 Massachusetts Institute of Technology, Cambridge, MA 02139;
 aml@space.mit.edu.}

\altaffiltext{4}{Instituto de Astrof\'{i}sica de Canarias, La Laguna,
Tenerife, Spain.}

\altaffiltext{5}{Department of 
Physics and Institute of Astronomy, National Tsing Hua University, 
Hsinchu, Taiwan.}

\altaffiltext{6}{South African Astronomical Observatory,
P.O. Box 9, Observatory, 7935, South Africa.}




\begin{abstract}

Results of observations of the Galactic bulge X-ray source GX 9+9 by
the All-Sky Monitor (ASM) and Proportional Counter Array (PCA) onboard
the \textit{Rossi X-ray Timing Explorer} are presented.  The ASM
results show that the 4.19 hour X-ray periodicity first reported by
Hertz and Wood in 1987 was weak or not detected for most of the
mission prior to late 2004, but then became strong and remained strong
for approximately 2 years after which it weakened considerably.  When
the modulation at the 4.19 hour period is strong, it appears in folded
light curves as an intensity dip over $\lesssim30$\% of a cycle and is
distinctly nonsinusoidal.  A number of PCA observations of GX 9+9 were
performed before the appearance of strong modulation; two were
performed in 2006 during the epoch of strong modulation.  Data
obtained from the earlier PCA observations yield at best limited
evidence of the presence of phase-dependent intensity changes, while
the data from the later observations confirm the presence of flux
minima with depths and phases compatible with those apparent in folded
ASM light curves.  Light curves from a \textit{Chandra} observation of
GX 9+9 performed in the year 2000 prior to the start of strong
modulation show the possible presence of shallow dips at the predicted
times.  Optical observations performed in 2006 while the X-ray
modulation was strong do not show an increase in the degree of
modulation at the 4.19 hour period.  Implications of the changes in
modulation strength in X-rays and other observational results are
considered.

\end{abstract}





\section{Introduction}

The bright Galactic X-ray source GX 9+9 has the characteristics of a
low-mass X-ray binary (LMXB) system and particularly of the atoll-type
LMXB subclass that is defined by the behavior of the source X-ray
energy and power density spectra
\citep{mayer70,schulz89,hasvdk89,2006csxs}.  A periodicity of $4.19
\pm 0.02$ hours in its X-ray flux was discovered using {\it HEAO} A-1
data by \citet{hw88}.  The periodic component was weak; its strength,
in terms of the amplitude of the best-fit sinusoid, was approximately
$4\%$ of the average flux.  \citeauthor{hw88} interpreted the
modulation as indicative of the orbital period of the binary system.
Under the assumption that the system consists of a 1.4~M$_\sun$
neutron star and a Roche-lobe-filling lower-main-sequence star, they
inferred that the latter is an early M dwarf with mass $M \approx$
0.2-0.45 M$_\sun$ and radius $R\approx$ 0.3-0.6 R$_\sun$.  They used
these values to put an upper bound on the orbital inclination of $i
\lesssim 63 ^\circ$ but, as noted by \citet{s90}, apparently they
calculated $R_2/a$ incorrectly; the correct bound on the inclination
is $i \lesssim 77 ^\circ$.

\citet{s90} found variations of the brightness of the optical
counterpart of GX 9+9 at essentially the same period, viz. $4.198 \pm
0.009$ hours.  The modulation amplitude was about 0.19 mag peak to
peak in the B band in 1987 and appeared to be somewhat larger in 1988.
In the latter observations, the amplitude was roughly the same in each
of the $B, V,$ and $R$ bands.

\citet{kong06} carried out nearly simultaneous observations of GX 9+9
in 1999 in the optical at the Radcliffe Telescope of the South African
Astronomical Observatory and in X-rays with the Proportional Counter
Array (PCA) instrument on the {\it Rossi X-ray Timing Explorer} ({\it
RXTE}). They found that modulation at the presumed orbital period was
clearly present in the optical, but no modulation at that period was
apparent in either the 2-3.5 keV or the 9.7-16 keV photon energy
bands. \citeauthor{kong06} also reviewed archival data from a 14.4
hour observation of GX 9+9 made with \textit{EXOSAT} in 1983, a 6.6
hour observation made with \textit{ASCA} in 1994, and a 6.4 hour
observation made with \textit{BeppoSAX} in 2000, but did not find
significant evidence of the 4.19 hour periodicity.
 
\citet{corn_etal07} obtained phase-resolved spectra of the optical
counterpart in 2004 May and found emission lines which comprise
components with differing radial velocity variations that must
originate in different places in the binary system, one of which is
likely to be the X-ray-illuminated face of the secondary.  The 4.19
hour period is clearly evident in the radial velocity variations.
Thus, these observations strongly confirm that this is the orbital
period.  The phenomenon was most clear in the He II $\lambda$4686 line
but was also evident in the Bowen blend and around H$\beta$.
Following early reports of the detection of the 4.19-h period in the
{\it RXTE} All-Sky Monitor (ASM) data and variation of the strength of
the modulation \citep{lev06,levcor06}, \citet{corn_etal07} used the
ASM data to estimate the period and time of X-ray minimum.  They were
then able to relate the time of X-ray minimum, the phases of the
radial velocity variations, and the variation of the brightness of the
optical counterpart in the blue continuum to each other.  They also
derived model-dependent lower and upper limits on the binary mass
ratio and on the orbital velocity of the secondary.

Herein we present the evidence that the X-ray photometric signature of
the (presumed) binary orbit of GX~9+9 has undergone dramatic changes
over the $\sim$12 years that the source has been monitored with the
ASM.  In addition to describing the results of our analysis of ASM
data, we present results of analysis of a number of PCA observations,
of an observation of GX~9+9 performed in 2000 with the {\it Chandra
X-ray Observatory}, and of optical observations performed in 2006.  In
Section 2 we describe the instrumentation, observations, and the
results from the observations.  In Section 3, we summarize our
results, suggest for the first time in the literature, to our
knowledge, that the observed modulation is closely related to the
dipping seen in other LMXBs, and discuss in general the possible
implications of this investigation.


\section{Instrumentation, Observations, and Results}

\subsection{ASM Observations and Results}

The ASM consists of three Scanning Shadow Cameras (SSCs) mounted on a
rotating Drive Assembly \citep{asm96}.  Approximately 50,000
measurements of the intensity of GX 9+9, each from a 90-s exposure
with a single SSC, were obtained from the beginning of the {\it RXTE}
mission in early 1996 through 2007.  A single exposure yields
intensity estimates in each of three spectral bands which nominally
correspond to photon energy ranges of 1.5-3, 3-5, and 5-12 keV with
sensitivity of a few SSC counts s$^{-1}$ (the Crab Nebula produces
intensities of 27, 23, and 25 SSC counts s$^{-1}$ in the 3 bands,
respectively).  For the analyses presented herein, we use observation
times that have been adjusted so as to represent times at the
barycenter of the Solar System.

\begin{figure*}[bth]
\centering
\includegraphics[height=6.0in,angle=270.0]{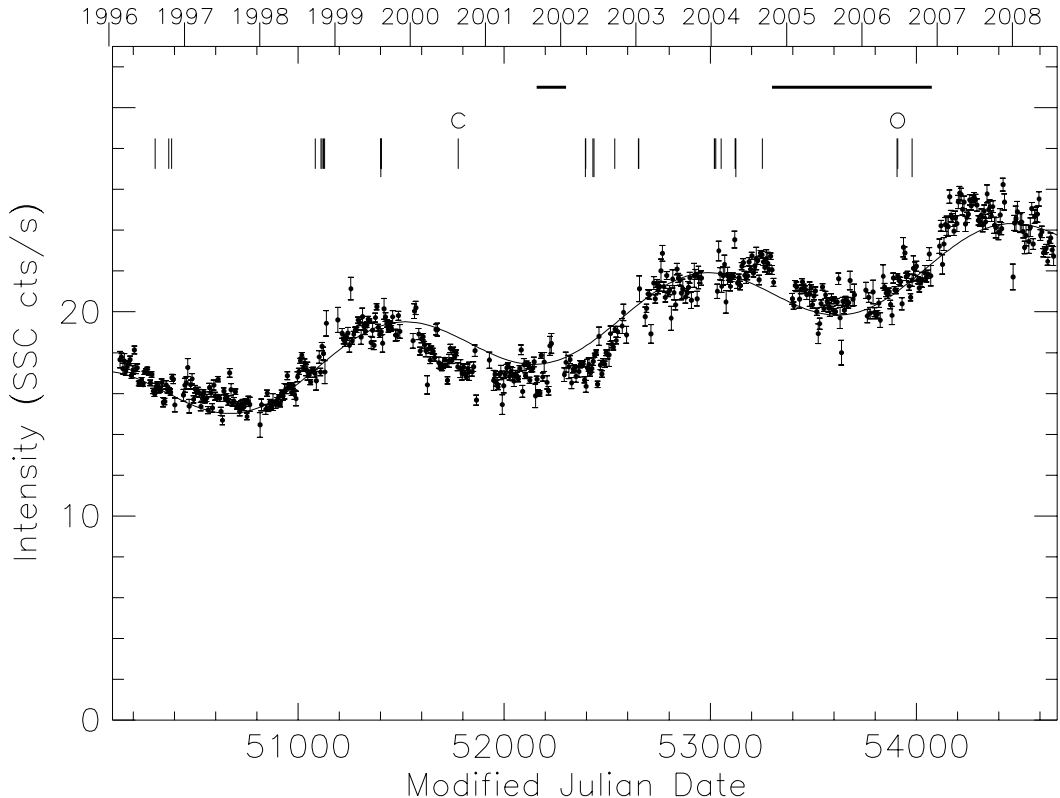}
\caption{ASM 1.5-12 keV light curve of GX 9+9 with measurements
averaged in contiguous 7 day time bins.  For times earlier than MJD
54200, the measurements include data from all three SSCs.  After MJD
54200, this light curve is based only on data from SSCs 2 and 3. Those
weighted average measurements with estimated uncertainties larger than
0.65 SSC counts s$^{-1}$ are not shown.  For reference, the Crab
Nebula intensity in the same band corresponds to 75 SSC counts
s$^{-1}$. The smooth curve is the best-fit function consisting of a
constant, linear term, and sinusoid (see text).  The period of the
sinusoid is 1473 days.  The vertical indicator marks show the times of
various observations. The marks without letters show the times of the
29 PCA observations of GX~9+9 that were longer than 1 hour in
duration; the longer marks show the times of the PCA observations with
exposures greater than 4 hours.  The time of the {Chandra X-ray
Observatory} observation is denoted with a ``C''; the time of the
optical observations reported herein are denoted with an ``O''.  The
heavy horizontal indicator bars approximately mark the times when the
X-ray modulation was easily detectable.
\label{fig:lc}}
\end{figure*}
The properties of the three SSCs have evolved over the $\sim 12$ years
of operation of the ASM.  SSC 1 (of SSCs 1 - 3) has a slow gas leak
that has resulted in the photon energy to pulse height conversion gain
increasing by about 10\% yr$^{-1}$ or about a factor of three over 12
years.  A number of the proportional counter cells in SSCs 2 and 3
have become permanently inoperational at various times since the
beginning of the mission because of catastrophic high voltage
breakdown events that removed the thin carbon layer on the
carbon-coated quartz fiber anodes of those cells.  In all three
detectors the properties of the carbon coatings of the anodes have
gradually changed.  These effects are, to first order, removed in the
course of the standard analysis procedure of the raw ASM data by the
ASM team that produces the light curve files.  The procedure is
adjusted so that the observed intensities of the Crab Nebula in the
four energy bands are more or less stable and close to the intensities
seen in SSC 1 in March 1996. The results in the 5-12 keV energy band
from SSC 1 at late times are corrected by the largest factors since
most events due to photons in this energy range in this SSC are not
usable because they are saturated in the amplifiers.  Thus, in
preparing light curves from GX9+9 for this paper, we compared the
results from SSC 1 with those from SSCs 2 and 3.  We found that the
SSC 1 light curve deviated negligibly from that produced from SSCs 2
and 3 before MJD 54200 and after that time gradually increased up to
$\sim 1$ SSC ct s$^{-1}$ above the intensities derived from SSCs 2 and
3 by MJD 54600. 


The ASM light curve of GX~9+9 is shown in Figure \ref{fig:lc}.  It was
made using data from all three SSCs for times prior to MJD 54200 and
from only SSCs 2 and 3 for times after MJD 54200.  It shows a
relatively strong source that varies on time scales of years.  The
intensity changes include a distinct sinusoidal-like component
superimposed on a slowly increasing baseline.  No spectral changes are
apparent in the ASM light curves beyond small differences that are
likely to be of instrumental origin.


To obtain a quantitative description of the long-period
sinusoidal-like component, we fit the light curve shown in Figure
\ref{fig:lc} with the simple function
\begin{equation}
F_x = a + b(t-50100) + c \sin(2\pi t/P_{\rm long}) + d \cos(2\pi t/P_{\rm long})
\end{equation}
where $F_x$ is the model 1.5-12 keV X-ray intensity, $t$ is the time
as a Modified Julian Date, $P_{\rm long}$ is the to-be-determined
period of the sinusoid, and $a, b, c$ and $d$ are also parameters to
be determined.  The best fit curve, with $P_{\rm long} = 1473$ days,
is shown superimposed on the ASM light curve in Figure \ref{fig:lc}.
The fit is not good, i.e., the value of the reduced $\chi^2$ statistic
is much greater than one.  Given this fit quality and the small number
of cycles of the sinusoid, we cannot confidently conclude that the
long-term light curve contains a periodic component that is coherent
over more than $\sim$3 cycles. Nonetheless, the form of the light
curve suggests that some type of quasiperiodic long-term variability
may be present.

In the early stages of a search for periodicities in ASM data using
advanced analysis techniques, \citet{shiv05} applied the first of two
primary strategies of improving the sensitivity of the search to high
frequency variations, and detected the 4.19 hour periodicity in the
ASM light curve of GX 9+9.  The first strategy involves the use of
appropriate weights such as the reciprocals of the variances in
Fourier and other types of analysis since the individual ASM
measurements have a wide range of associated uncertainties.  The
second strategy stems from the fact that the observations of the
source are obtained with a low duty cycle, i.e., the window function
is sparse (and complex).  The properties of the window function, in
combination with the presence of slow variations of the source
intensity, act to hinder the detection of variations on short time
scales.  The window function power density spectrum (PDS) has
substantial power at high frequencies, e.g., 1 cycle d$^{-1}$ and 1
cycle per spacecraft orbit ($\sim95$ minute period).  Since the data
may be regarded as the product of a (hypothetical) continuous set of
source intensity measurements with the window function, a Fourier
transform of the data is equivalent to the convolution of a transform
of a continuous set of intensity measurements with the window function
transform.  The high frequency structure in the window function
transform acts to spread power at low frequencies in the source
intensity to high frequencies in the calculated transform (or,
equivalently, the PDS).  This effectively raises the noise level at
high frequencies.

\begin{figure}[tbh]
\centering
\includegraphics[height=3.4in,angle=90.0]{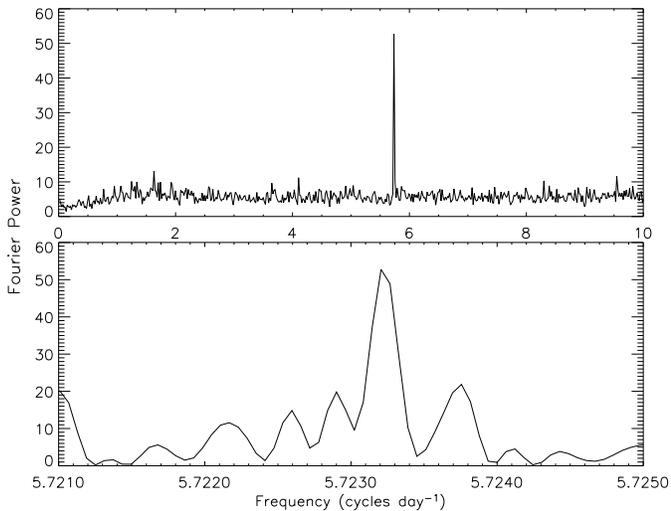}
\caption{Power density spectrum (PDS) of a 1.5-12 keV ASM light curve of
GX 9+9 that had been modified to remove variability on time scales
longer than $\sim1$ day (see text).  The original FFT was oversampled
and had approximately 2 million frequencies from 0 cycles day$^{-1}$
to the Nyquist frequency of 144 cycles day$^{-1}$.  (upper panel) A
rebinned PDS in which the number of frequency bins was reduced by a
factor of 300 by using the maximum power in each contiguous set of 300
frequency bins in the original PDS as the power of the corresponding
bin in the rebinned PDS.  (lower panel) A portion of the original
PDS. In both panels, the power is normalized relative to the PDS-wide
average.
\label{fig:pds}}
\end{figure}

In our analysis, the sensitivity to high frequency variations is
enhanced by subtracting a smoothed version of the light curve from the
unsmoothed light curve.  To perform the smoothing, we do not simply
convolve a box function with the binned light curve data since that
would not yield any improvement in the noise level at high
frequencies.  Rather, we ignore bins which do not contain any actual
measurements and we use weights based on estimates of the
uncertainties in the individual measurements to compute the smoothed
light curve.  The ``box'' used in the smoothing had a length of 0.9
day, so the smoothed light curve displayed only that variability with
Fourier components at frequencies below $\sim 1$ day$^{-1}$. The
smoothed light curve was subtracted from the unsmoothed light curve,
and the difference light curve was Fourier transformed.  The results
are illustrated in Figure \ref{fig:pds}.  The center frequency and
width (half-width at half-maximum) of the peak correspond to a period
of $4.19344 \pm 0.00007$ hours ($0.1747267 \pm 0.0000029$ days).

\begin{figure*}[tbh]
\centering
\includegraphics[height=6.0in]{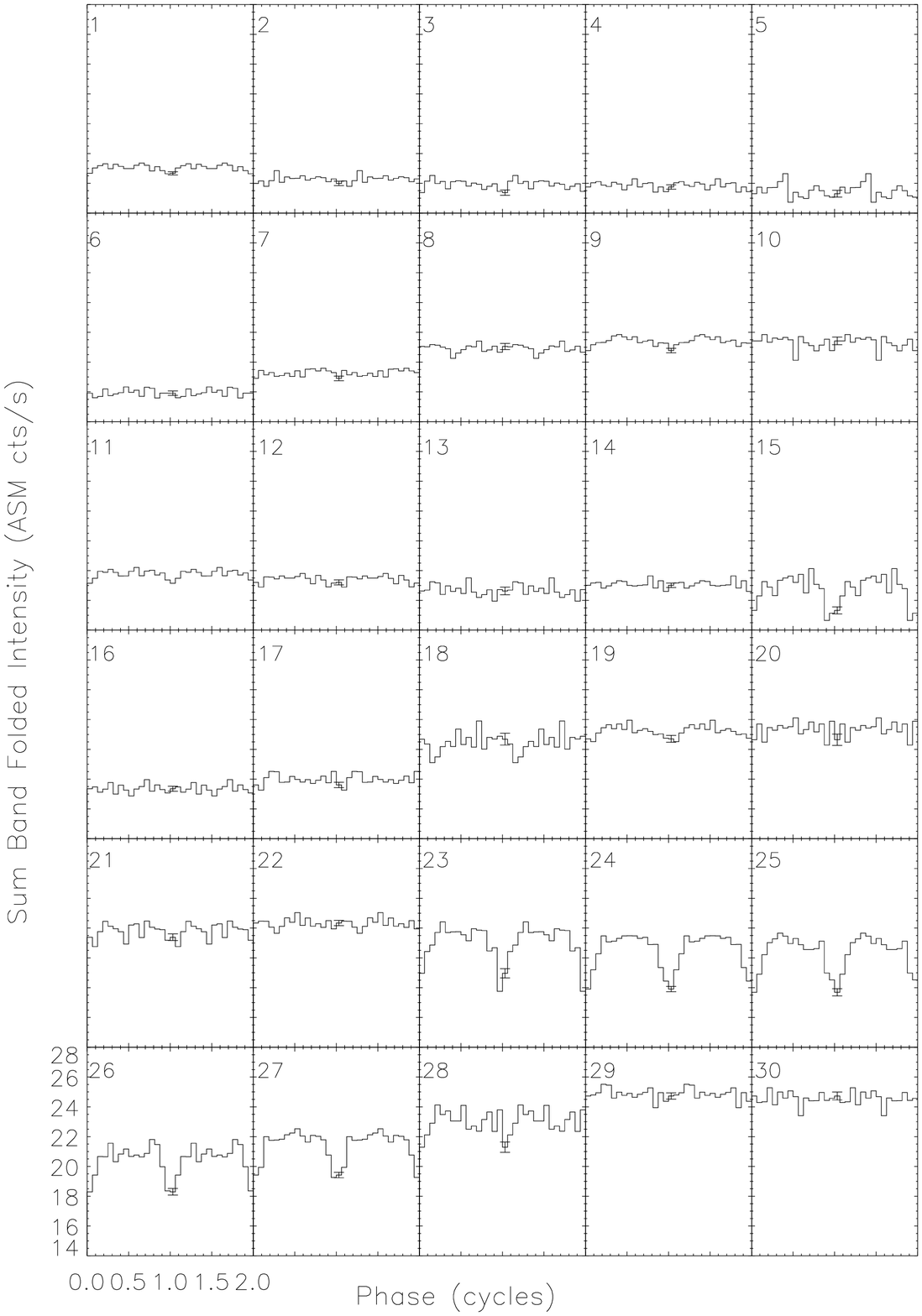}
\caption{Barycenter-corrected ASM 1.5-12 keV light curves of GX 9+9 from
each of 30 equal-duration time intervals folded using the epoch and
period in eq. (2), i.e., MJD 53382.959 and 4.19344 h = 0.1747267 d,
respectively.  Each interval spans $\sim144.40$ days; dates may be
found in Table \ref{tbl-1}.  All vertical axes have the same scale and
offset, as do all horizontal axes.  The interval numbers are given in
the top left corners. Typical $\pm 1 \sigma$ uncertainties are shown.
\label{fig:lcpcs}}
\end{figure*}

To investigate the time variability of the orbital modulation, we
folded the barycenter-corrected light curves for each of 30 equal time
intervals at the 4.19344 hour period; see Figure \ref{fig:lcpcs}.  The
figure shows that the modulation of GX 9+9 was weak during the first 5
years of the mission.  It then seems to have appeared at a somewhat
detectable level for a brief time in late 2001 and/or early 2002
(panel 15).  In late 2004 or early 2005 the modulation grew stronger,
and it stayed strong until late in 2006 when it began to decrease in
strength. In the final two panels the modulation is not present at a
significant level.  We note that the interval of relatively strong
modulation, i.e., approximately from MJD 53300 to MJD 54100,
corresponds to a time interval in which the overall intensity is below
the levels given by an interpolation from earlier to later times (see
Fig.~\ref{fig:lc}).

\begin{deluxetable}{lll}
\tablewidth{0pt}
\tabletypesize{\footnotesize}
\tablecaption{Time Intervals of Figure \ref{fig:lcpcs} Panels\label{tbl-1}}
\tablehead{
\colhead{Interval}      & \colhead{Begin MJD\tablenotemark{a}} &
\colhead{Begin Date\tablenotemark{b}} }
\startdata
1 & 50135.0 & 1996 Feb 22 \\
2 & 50279.4 & 1996 Jul 15 \\
3 & 50423.8 & 1996 Dec 6 \\
4 & 50568.2 & 1997 Apr 30 \\
5 & 50712.6 & 1997 Sep 21 \\
6 & 50857.0 & 1998 Feb 13 \\
7 & 51001.4 & 1998 Jul 7 \\
8 & 51145.8 & 1998 Nov 28 \\
9 & 51290.2 & 1999 Apr 22 \\
10 & 51434.6 & 1999 Sep 13 \\
11 & 51579.0 & 2000 Feb 5 \\
12 & 51723.4 & 2000 Jun 28 \\
13 & 51867.8 & 2000 Nov 19 \\
14 & 52012.2 & 2001 Apr 13 \\
15 & 52156.6 & 2001 Sep 4 \\
16 & 52301.0 & 2002 Jan 27 \\
17 & 52445.4 & 2002 Jun 20 \\
18 & 52589.8 & 2002 Nov 11 \\
19 & 52734.2 & 2003 Apr 5 \\
20 & 52878.6 & 2003 Aug 27 \\
21 & 53023.0 & 2004 Jan 19 \\
22 & 53167.4 & 2004 Jun 11 \\
23 & 53311.8 & 2004 Nov 2 \\
24 & 53456.2 & 2005 Mar 27 \\
25 & 53600.6 & 2005 Aug 18 \\
26 & 53745.0 & 2006 Jan 10 \\
27 & 53889.4 & 2006 Jun 3 \\
28 & 54033.8 & 2006 Oct 25 \\
29 & 54178.2 & 2007 Mar 19 \\
30 & 54322.6 & 2007 Aug 10 \\
31\tablenotemark{c} & 54467.0 & 2008 Jan 2
\enddata
\tablenotetext{a}{Modified Julian Date rounded to one decimal place.}
\tablenotetext{b}{Corresponding to the rounded MJD.}
\tablenotetext{c}{The end of interval 30.}
\end{deluxetable}

\begin{figure*}[tbh]
\epsscale{0.7}
\plotone{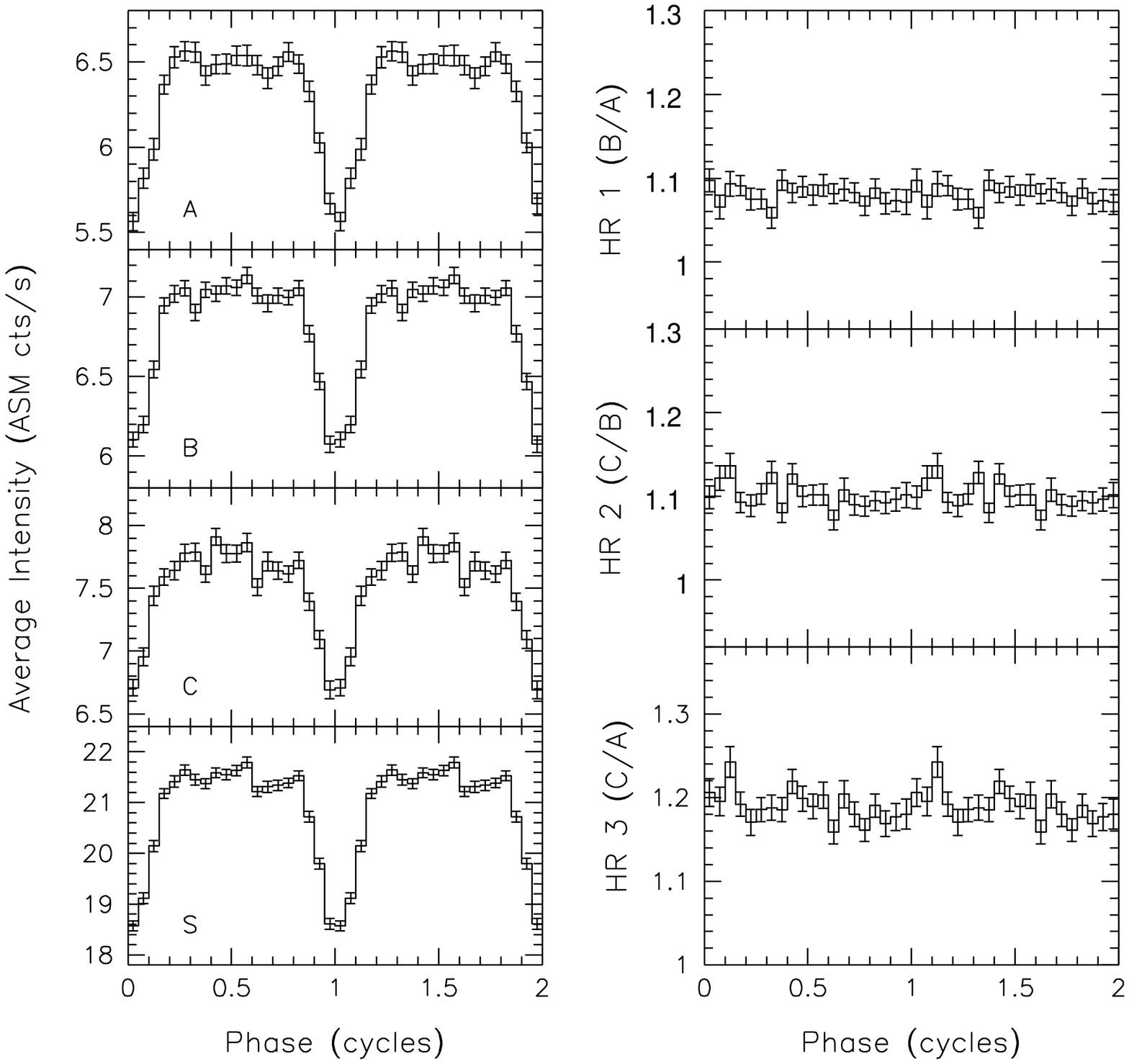}
\caption{(left) Folded barycenter-corrected light curves from ASM
observations of GX 9+9 from 2004 October 22 (MJD 53300) through 2006
December 6 (MJD 54075).  The energy bands are 1.5-3 keV (A), 3-5 keV
(B), 5-12 keV (C), and 1.5-12 keV (S). The folding was accomplished
using the period and epoch in eq. (2). (right) Hardness ratios
implied by bin-wise ratios of folded intensities.
\label{fig:latefold}}
\end{figure*}

In Figure \ref{fig:latefold} we show folded light curves for the
1.5-3, 3-5, and 5-12 keV photon-energy bands as well as for the
overall 1.5-12 keV band for the time interval of MJD 53300 through MJD
54075 when the modulation was strongest. The modulation appears to be
largely energy-independent and, in particular, the spectrum of GX 9+9
does not significantly change during the dips.  This is confirmed by
the hardness ratios computed from these folded light curves by taking
bin-wise ratios as shown in the right-hand side of
Fig.~\ref{fig:latefold}.

We have used the folded light curves to determine the epoch of X-ray
minimum, i.e., the time corresponding to the centroid of the dip-like
feature.  The best linear fit to the times of the minima is given by:
\begin{equation}
T_{\rm min} = 53382.959 \pm 0.003 + (0.1747267 \pm 0.0000029)N
\end{equation}
where the times are Modified Julian Dates in the UTC time system and
apply at the barycenter of the Solar System.  One should note that,
per Fig.~\ref{fig:latefold}, dip activity may occur within $\pm0.15$
cycles ($= \pm0.63$ hours) of the times of minimum.  It must be
pointed out that this ephemeris is not consistent with that of
\citet{corn_etal07}; their time of minimum occurs at $0.24 \pm 0.04$
cycles relative to a time of minimum given by this equation.

\subsection{PCA Observations and Results}

The PCA consists of 5 mechanically-collimated large area Proportional
Counter Units (PCUs) and is primarily sensitive to 2-60 keV photons
\citep[][and references therein]{pca06}.  A total of 61 observations
of GX 9+9 were carried out with the PCA from 1996 through 2004.  Only
29 of these were over one hour in duration; 17 were over 4 hours; the
longest was $\sim 15$ hours.  Twelve of the 29 observations were
performed in the time period 1996-1999; 17 were performed between 2002
and 2004.  In addition, per our request a pointed observation of GX
9+9 was carried out on June 20, 2006.  One further observation of the
source was carried out on September 1, 2006. GX 9+9 typically produced
500 to 700 cts s$^{-1}$ PCU$^{-1}$ in the 2-12 keV energy band in
these observations.

For our periodicity search and X-ray ``color'' analyses using PCA data
we have used data accumulated in Standard Mode 2 which provides counts
in each of 128 pulse height channels for each PCU for 16 s time
intervals.  Data accumulated in various event and single-bit modes
that provided a time resolution of 500 $\mu$s or better were used for
our fast-timing analyses.  Background has been neglected for most of
the timing analyses. However, in the analyses involving X-ray colors,
the data have been corrected for both the background as well as gain
drifts over the course of the mission. The times of the data have been
corrected to the Solar System barycenter for all of the analyses
presented herein.

In response to the discovery of the strengthening of the orbital
modulation in the ASM data \citep{lev06}, we searched the pre-2006
archival PCA data for evidence of the 4.19 h periodicity.  In our
search we used only the 29 observations that were longer than 1 hour.
The total exposure obtained during these observations was $\sim460$
ks.  Since the ASM data show no significant modulation during the
times of the observations, we expected to find no evidence of
modulation in the PCA data.

We folded the barycenter-corrected data from the 29 observations at
each of a grid of closely spaced periods centered on 4.19344 hrs.  We
did this for the energy bands 2-5, 5-12, and 12-30 keV.  For each
folded light curve we computed a $\chi^2$ statistic based on the
hypothesis of no variation as a function of phase and plotted the
resulting values against folding period.  No significant peaks were
found.  We also found no significant changes in the color of the
source that correlated with the phase of the folds.

This search for periodic modulation was done against a background of
variability which tends to be stronger at higher than at lower photon
energies.  On timescales of 256 sec, the RMS variability seen in the
PCA light curves, expressed as a fraction of the mean count rate, is
$3.1 \%$ in the 2-6 keV band, $6.4 \%$ in the 6-10 keV band, and $9.6
\%$ in the 10-16 keV band.  On timescales of 24 min, the RMS is $3.3
\%$ in the 2-6 keV band, $5.1 \% $ in the 6-10 keV band, and $6.9 \%$
in the 10-16 keV band.  These results agree roughly with the ASM light
curves, which yield a $\sim 4 \%$ RMS intrinsic variability in each of
the A, B, and C bands.  This should be compared to the original
variability estimate of \citet{hw88} who calculated a $4.6 \%$
intrinsic source variability in addition to the $3.8 \%$ sinusoidal
modulation at the orbital period (4.19 hrs) in the 0.5-20 keV {\it
HEAO} A-1 lightcurve.

\begin{figure*}[tbh]
\centering
\includegraphics[angle=90,height=3.8in]{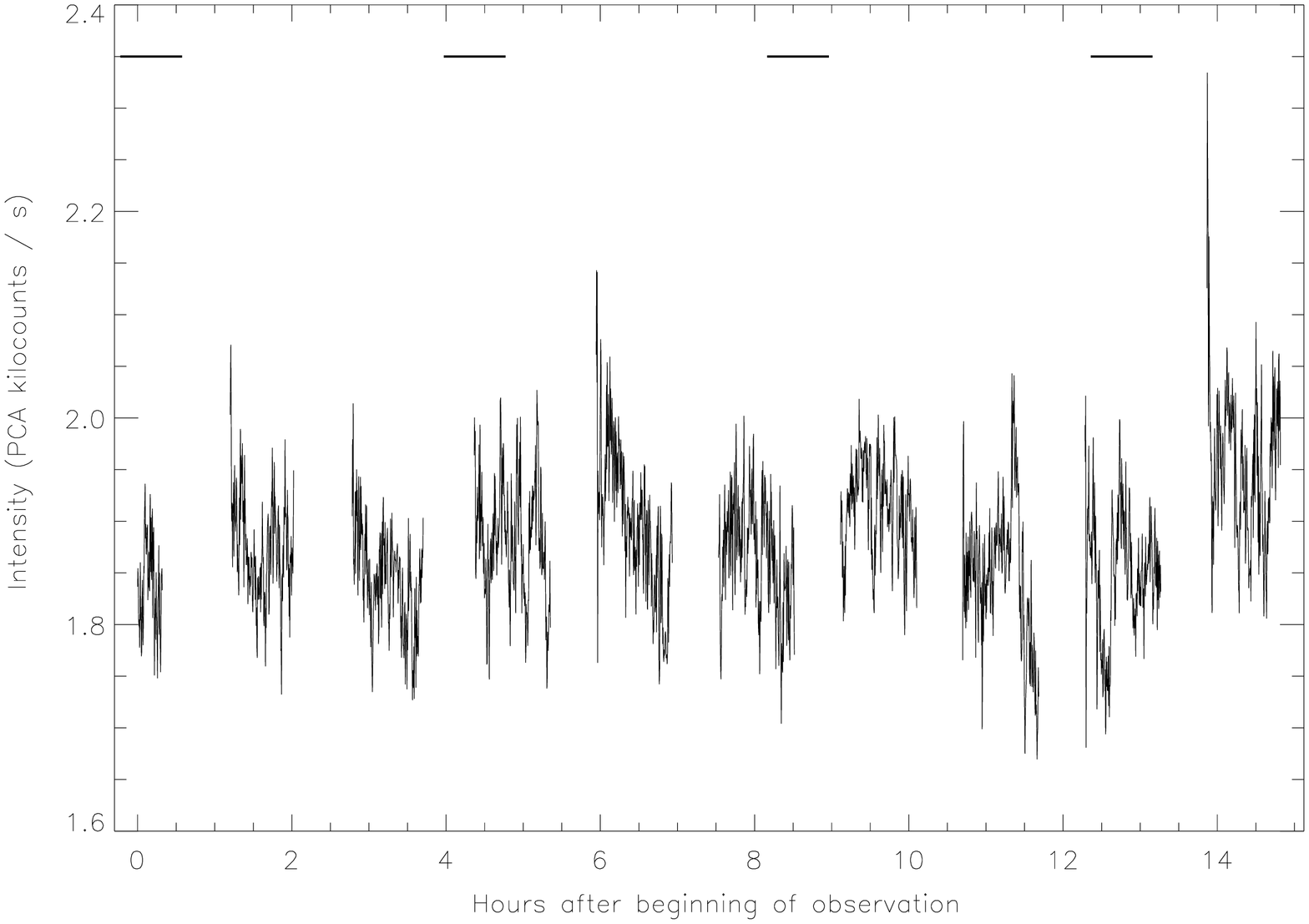}
\caption{Light curve of counts in the 2-6 keV band of GX 9+9
obtained from a PCA observation beginning at 13:42:20 UTC on 2002 May
1 (MJD 52395; observation ID 70022-02-01-01).  The horizontal bars
indicate the times of X-ray minimum according to eq. (2).  Note that
there is one possible dip at $\sim$ 8.3 hours which is consistent
with the phase of the ASM dip and there is another more noticeable dip
at 12.5 hours.
\label{fig:pcamay02}}
\end{figure*}

\begin{figure}[tbh]
\centering
\includegraphics[angle=0,height=3.5in]{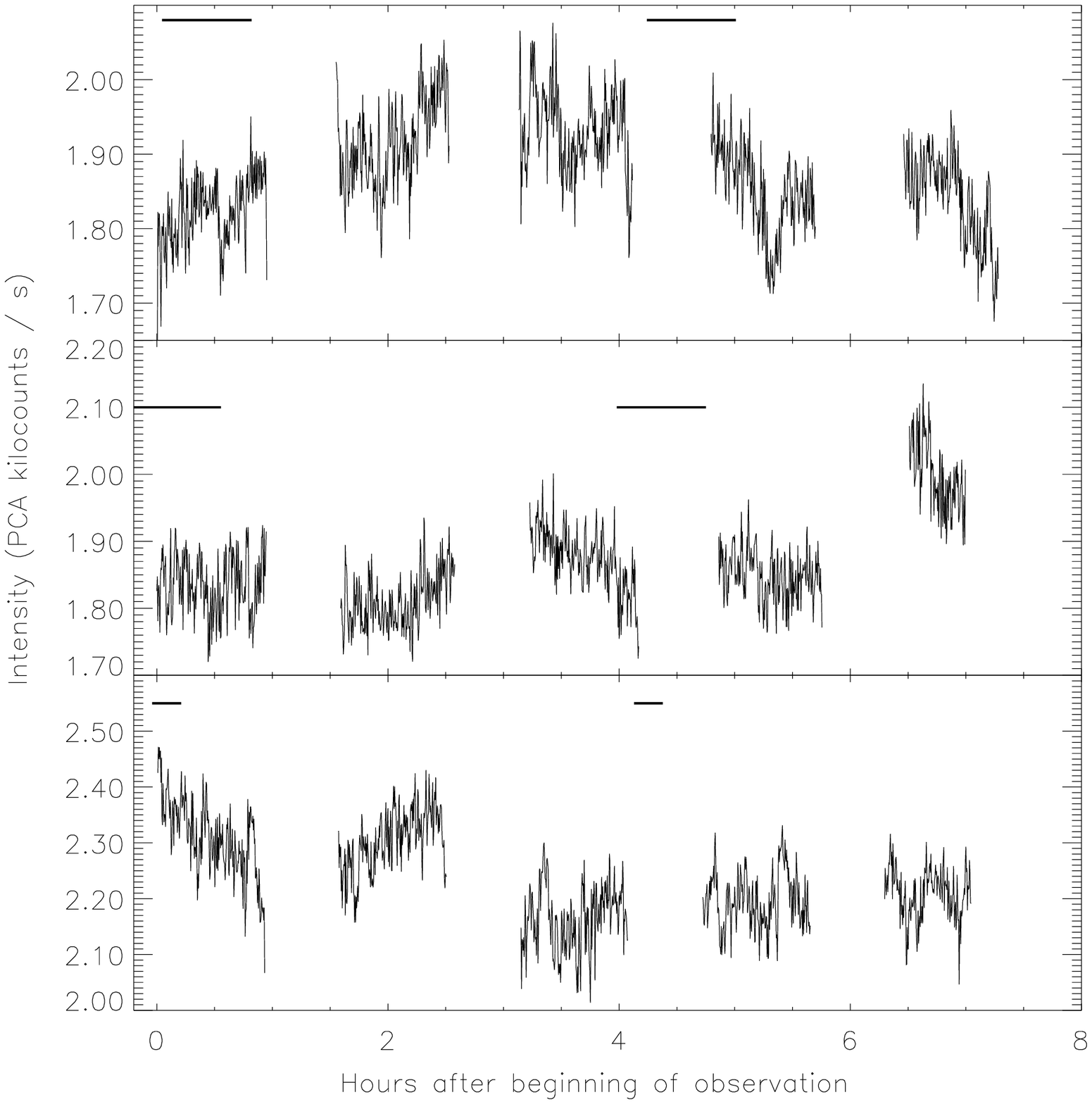}
\caption{PCA count rates in the 2-6 keV band during 3 observations of
GX 9+9.  (top) The observation beginning at 9:06:20 on 2002 June 6
(MJD 52431; observation ID 70022-02-05-00).  (middle) That beginning
at 10:58:36 on 2002 June 11 (MJD 52436; ID 70022-02-06-00.  (bottom)
That beginning at 17:03:08 on 2004 April 29 (MJD 53124; ID
70022-02-09-00).  The horizontal bars indicate the times of X-ray
minimum according to eq. (2).
\label{fig:pca3other}}
\end{figure}

While the folding analysis did not yield any evidence for persistent
modulation that was stable in phase, upon examination of the
individual pre-2006 PCA observations, dip-like events were evident on
rare occasions; they are illustrated in Figures \ref{fig:pcamay02} and
\ref{fig:pca3other}.  In the light curve from the observation on 2002
May 1 (Fig.~\ref{fig:pcamay02}), the most well-defined dip-like event
is seen near 12.5 hours. A weaker dip-like event is seen near 8.3
hours; it is not much more prominent than some other dip-like events
in this observation.  In the light curve from the observation on 2002
June 6 (top panel of Fig.~\ref{fig:pca3other}), narrow dips are
evident at about 0.6 and 5.3 hours. The times, orbital phases, and
depths of these dips are given in Table~\ref{tbl-2}.  Prominent dips
are not seen in the light curves from 2002 June 11, merely 5 days
later, nor are they evident in the light curves from 2004 April 29.
The dips seen on 2002 May 1 and on 2002 June 6 are shorter than the
dips seen in the folded ASM light curves.  The time of one of these
dips is not centered on a time given by eq. (2), but the phases of all
four may be consistent with the extent in phase of the dips seen in
the folded ASM light curves.  We did not find any other particularly
noticeable dips in the light curves from any of the other pre-2006 PCA
observations that were one or more hours in duration.

\begin{figure*}[tbh]
\centering
\epsscale{1.08}
\plottwo{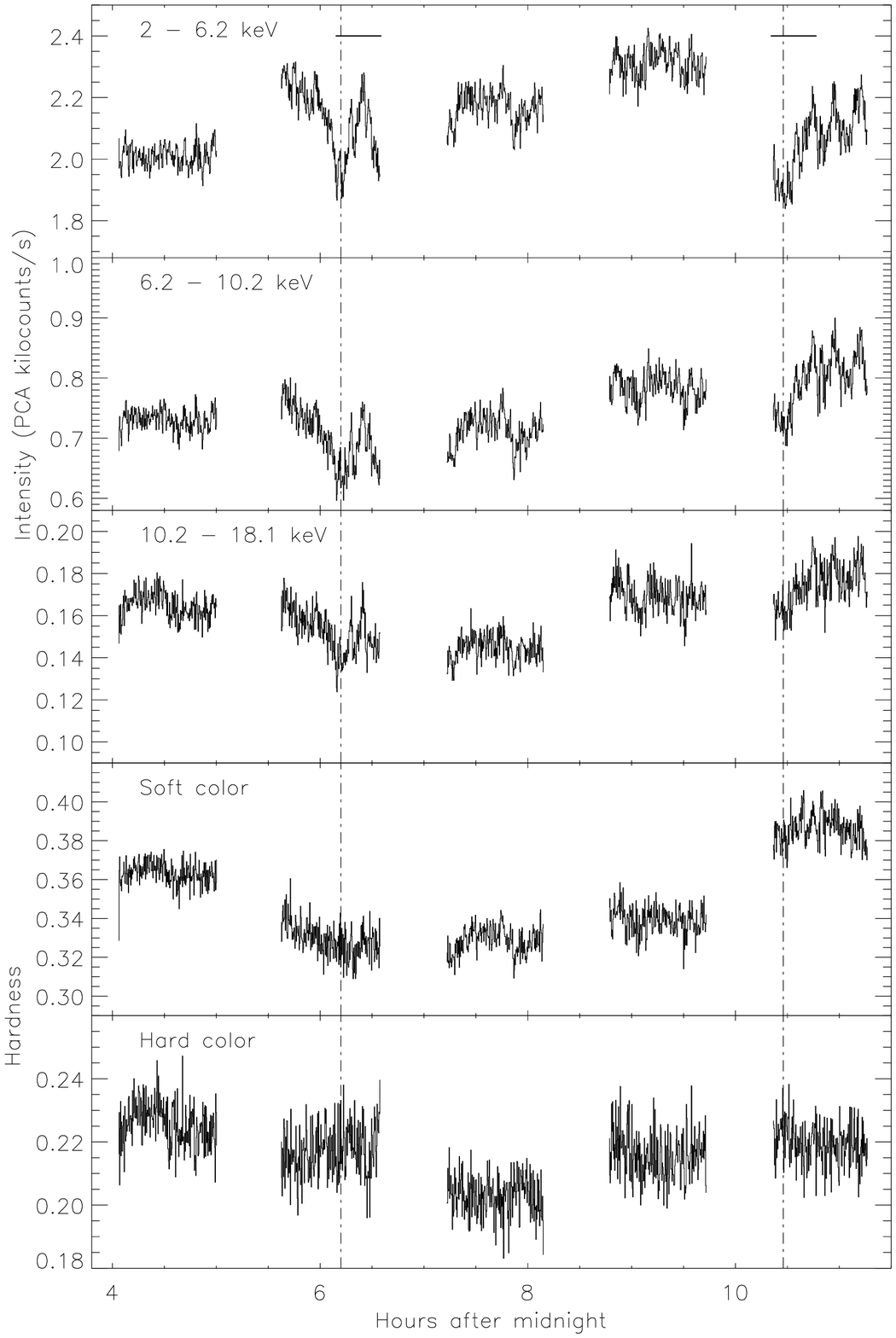}{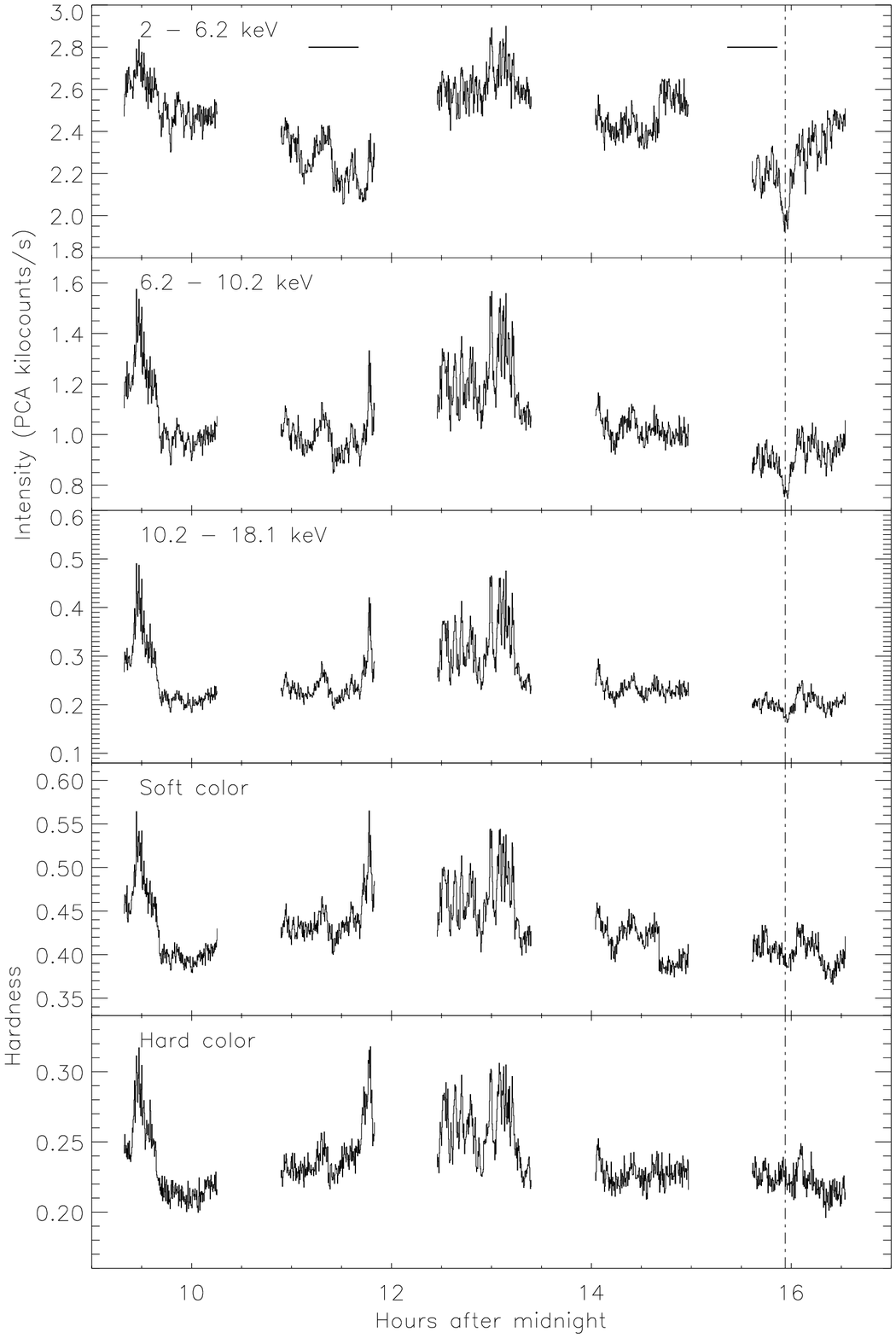}
\caption{(left) PCA Observation of GX 9+9 taken on June 20, 2006 (MJD
53906; ID 92415-01-01-00).  (right) PCA Observation of GX 9+9 taken on
September 1, 2006 (MJD 53979; ID 92415-01-02-00).  Both time axes are
labelled in hours after midnight TT of the day of the observation.
The horizontal bars indicate the times of X-ray minimum according to
eq. (2).  The panels labelled ``soft color'' and ``hard color'' are,
respectively, the ratio of counts in the 6.2-10.2 keV band to that in
the 2-6.2 keV band, and the ratio of counts in the 10.2-18.1 keV band
to that in the 6.2-10.2 keV band.  The vertical dash-dot lines mark
the center times of notable dip-like events.
\label{fig:pcajune06}}
\end{figure*}

Per our request to the {\it RXTE} Mission Scientist for new
observations, GX~9+9 was observed for 10 h with the PCA (and HEXTE) on
2006 June 20 (MJD 53906).  The results are shown in three different
energy bands in Figure \ref{fig:pcajune06}.  According to eq. (2); we
find that the times of minimum flux should occur at $6.37 \pm 0.22$
hours and $10.56 \pm 0.22$ hours (TT) on 2006 June 20.  Indeed,
prominent dips in the intensity can be seen close to these predicted
times in the two lower energy bands and at the earlier of the two
times in the 10-18 keV band.  Estimates of the times, orbital phases,
and depths of these two dips are given in Table~\ref{tbl-2} wherein
the depths are seen to be more or less independent of energy.  It is
not clear whether the second of these two dips is less prominent in
the 10-18 keV band (Fig.~\ref{fig:pcajune06}) because of generally
stronger variability at higher than at lower energies or,
alternatively, that it actually is less deep and the uncertainty on
the depth given in Table~\ref{tbl-2} is underestimated.

Another pointed observation was done on 2006 September 1 (MJD 53979).
Like that on 2006 June 20, this one also took place during an interval
(Fig.~\ref{fig:lcpcs}) in which the source exhibited strong
modulation.  With respect to the time reference used for the right
half of Fig.~\ref{fig:pcajune06}, eq. (2) predicts times of minimum
flux at $11.42 \pm 0.25$ and $15.61 \pm 0.25$ hours.  In this
observation, dips are evident near 11.7 and 15.9 hours.  They are most
apparent in the 2-6 keV band.  The first dip is hardly evident in the
6-10 and 10-18 keV bands, perhaps because of a flare in which the
spectral hardness undergoes a large increase.  The dip near 15.9 hours
is quite evident in both higher energy bands; times, phases, and
fractional depths are given in Table~\ref{tbl-2}. The depths of this
dip are, like those seen on 2006 June 20, also more or less
independent of energy.  We have not determined the depths or phase of
the dip near 11.7 hours because of its proximity to the flare.

The two dips seen in the 2006 June observation and the second dip seen
in the 2006 September observation are comparable in terms of
fractional depth with those seen in the folded ASM light curves (see
interval 27 in Fig. \ref{fig:lcpcs}), and are, as noted immediately
above, energy independent to a good approximation.  They are narrower
($\sim 0.05$ orbital cycles) than the dips in the folded ASM light
curves.  The phases of the three dips fall within the extent in phase
of the dips in the folded ASM light curves.

\begin{deluxetable*}{ccccccc}
\tablewidth{0pt}
\tabletypesize{\footnotesize}
\tablecaption{Dip Times and Depths in PCA Data\label{tbl-2}}
\tablehead{
\colhead{Figure No.} &
\colhead{ObsID}      &
\colhead{Energy Band} &
\colhead{Centroid Time\tablenotemark{a}} &
\colhead{Centroid Time\tablenotemark{b}} &
\colhead{Orbital Phase\tablenotemark{c}} &
\colhead{Depth} \\
  &
  &
\colhead{(keV)} &
\colhead{(hours)} &
\colhead{(MJD)} &
\colhead{(cycles)} &
\colhead{(fraction)}
 }
\startdata
5 & 70022-02-01-01 & 2-6 &  $8.363 \pm 0.02$ & 52395.9195 & $-0.048 \pm 0.095$ & $0.031 \pm 0.004$ \\
5 & 70022-02-01-01 & 2-6 &  $12.548 \pm 0.01$ & 52396.0939 & $-0.050 \pm 0.095$ & $0.041 \pm 0.002$ \\
6 & 70022-02-05-00 & 2-6 &  $0.575 \pm 0.01$ & 52431.4034 & $0.034 \pm 0.092$ & $0.026 \pm 0.006$ \\
6 & 70022-02-05-00 & 2-6 &  $5.328 \pm 0.01$ & 52431.6014 & $0.168 \pm 0.092$ & $0.067 \pm 0.008$ \\
7(left) & 92415-01-01-00 & 2-6 & $6.201 \pm 0.01$ & 53906.2584 & $-0.041 \pm 0.053$ & $0.137 \pm 0.003$ \\
7(left) & 92415-01-01-00 & 6-10 & $6.212 \pm 0.01$ & 53906.2588 & $-0.038 \pm 0.053$ & $0.172 \pm 0.004$ \\
7(left) & 92415-01-01-00 & 10-18 & $6.204 \pm 0.02$ & 53906.2585 & $-0.040 \pm 0.053$ & $0.171 \pm 0.007$ \\
7(left) & 92415-01-01-00 & 2-6 & $10.461 \pm  0.02$ & 53906.4359 & $-0.025 \pm 0.053$ & $0.102 \pm 0.003$ \\
7(left) & 92415-01-01-00 & 6-10 & $10.461 \pm  0.01$ & 53906.4359 & $-0.025 \pm 0.053$ & $0.114 \pm 0.003$ \\
7(left) & 92415-01-01-00 & 10-18 & $10.443 \pm  0.05$ & 53906.4351 & $-0.029 \pm 0.054$ & $0.095 \pm 0.004$ \\
7(right) & 92415-01-02-00 & 2-6 & $15.937 \pm  0.01$ & 53979.6640 & $0.076 \pm 0.059$ & $0.098 \pm 0.007$ \\
7(right) & 92415-01-02-00 & 6-10 & $15.939 \pm  0.01$ & 53979.6641 & $0.077 \pm 0.059$ & $0.117 \pm 0.009$ \\
7(right) & 92415-01-02-00 & 10-18 & $15.940 \pm  0.01$ & 53979.6642 & $0.077 \pm 0.059$ & $0.108 \pm 0.013$
\enddata
\tablenotetext{a}{Time relative to the origin in the figure.}
\tablenotetext{b}{Centroid time as a Modified Julian Date.}
\tablenotetext{c}{Computed according to eq. (2).  The uncertainty is
the root-sum-squared of the 1-sigma errors in the projection of the
ephemeris given in eq. (2) and the time of the dip centroid.  The
uncertainties in the dip centroid times are small in comparison with
the ephemeris uncertainties.}
\end{deluxetable*}

After searching the PCA observations for evidence of persistent
modulation, we attempted to determine whether or not the
characteristics of the source's high frequency PDS had changed since
the increase of modulation.  We also attempted to see if any change in
the X-ray color related phenomenology had occurred in tandem with the
modulation strength increase.  We report on both the fast-timing and
color analyses below.

\subsection{Fast-Timing Analysis} 

Quasi-periodic oscillations (QPOs) are often seen in the X-ray
intensities of neutron-star LMXBs (see the review by
\citealt{2006csxs} and references therein).  Theoretical
considerations suggest that these oscillations are generated in the
accretion disks, although there is no consensus on the details of the
physical mechanisms causing the oscillations.  GX 9+9 has never been
reported to exhibit such QPOs \citep[see, e.g.,][]{wij98}.  In this
respect it is like the other bright atoll-type GX sources
\citep{2006csxs}.  Because of the discovery of the modulation-strength
increase in the source, we reviewed all available PCA observations,
but did not find statistically significant QPOs.  In particular, the
source did not exhibit statistically significant QPOs in the frequency
range $\sim 0.01 - 1000$ Hz during either the June 2006 or September
2006 observations.


\subsection{Color Analysis}

GX 9+9, like the other atoll-type GX sources, generally occupies the
upper and lower banana regions in a soft color - hard color diagram
\citep{schulz89,wij98,2006csxs}.  We wondered whether the presence of strong
modulation would have any impact on the details of the color-color
diagram.  We used data only from PCU 3 (of PCUs 1-5) for this
analysis.  The data were corrected for background and gain changes,
and then used to make color-color diagrams for both the June 2006 and
September 2006 observations.  We then compared these with a
color-color diagram made from all data from previous PCA observations
of GX 9+9.  We found no significant differences among the diagrams.


\subsection{Chandra Observations}

The \textit{Chandra X-ray Observatory} ({\it CXO}) comprises a high
angular resolution X-ray telescope and two focal plane cameras along
with a pair of transmission diffraction gratings each of which may be
placed in the X-ray beam just behind the mirror assembly for spectral
observations.  The ACIS-S focal plane array, which was used for the
observation described below, is composed of 4 front-illuminated (FI)
and 2 back-illuminated (BI) CCDs configured in a 6 by 1 array.  It
provides a field of approximately 8 arcminutes by 48 arcminutes.  The
on-axis effective area for either the FI or BI CCDs is 340 cm$^2$ at 1
keV, and the spectral resolution is $E/dE =$ 20-50 (1-6 keV) for the
FI CCDs and $E/dE =$ 9-35 (1-6 keV) for the BI CCDs.  See
\citet{chand02} for further information.

\begin{figure}[hbtp]
\centering
\includegraphics[height=3.3in,angle=0.0]{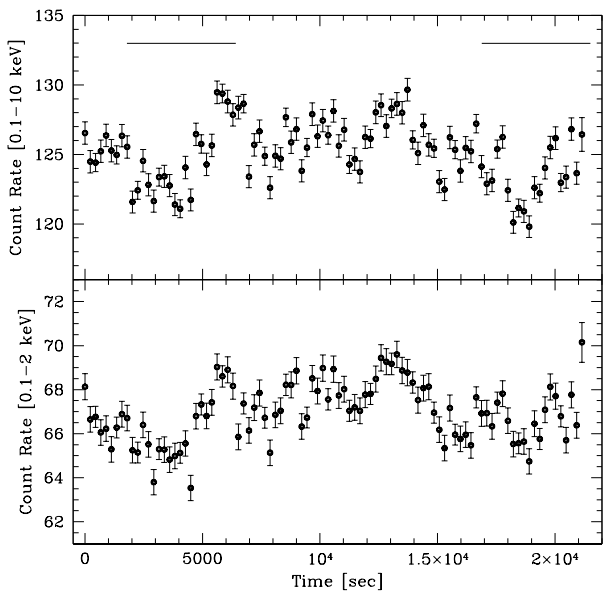}
\caption{Intensity as a function of time of GX~9+9 as shown by counts
in 200-s time bins obtained from the \textit{Chandra} ACIS-HETG
observation. Time is referenced to the beginning of the observation at
2000 August 22 05:20:21 (MJD 51778.2225). The horizontal bars indicate
the times of X-ray minimum according to eq. (2).
\label{fig:chandra}}
\end{figure}

A 20 ks long observation of GX 9+9 was carried out with the
Observatory on 2000 August 22 beginning at 05:20:21 UTC (MJD
51778.2225); the ACIS-S instrument was used as the focal plane camera,
and the High-Energy Transmission Grating (HETG) was in the X-ray beam.
The zero-order image of GX~9+9 was placed on the back-illuminated CCD
S3.  The count rate of good events in the dispersed spectra, i.e., not
including the counts in the zero-order image, is shown in
Figure~\ref{fig:chandra}.  Only small changes in the rate are
evident. Among these changes are two shallow minima with depths of
approximately 4\% of the mean count rate that are approximately 15 ks
apart.  These times are consistent with the times of X-ray minimum
according to eq. (2).

\subsection{Optical Observations}

We observed the field of GX 9+9 in June 2006 in white light, i.e.,
with no filter, with the CCD imaging camera on the 1.9\,m Radcliffe
Telescope of the South African Astronomical Observatory (SAAO). In
order to achieve rapid read-out, we only used a 127$\times$96\,pixel
data area, giving a read-out cycle of 6\,s. The observing run was
seven nights, of which two were clouded out and two partially affected
by poor weather. Seeing was variable, but under 1\arcsec\ for
substantial periods of the three good nights.

These observations were performed not long after the PCA observations
of GX 9+9 on 2006 June 20 while the orbital modulation in X-rays was
strong (see Figs.~\ref{fig:lcpcs} and \ref{fig:pcajune06}).

\begin{figure*}[tbh]
\centering
\epsscale{0.6}
\plotone{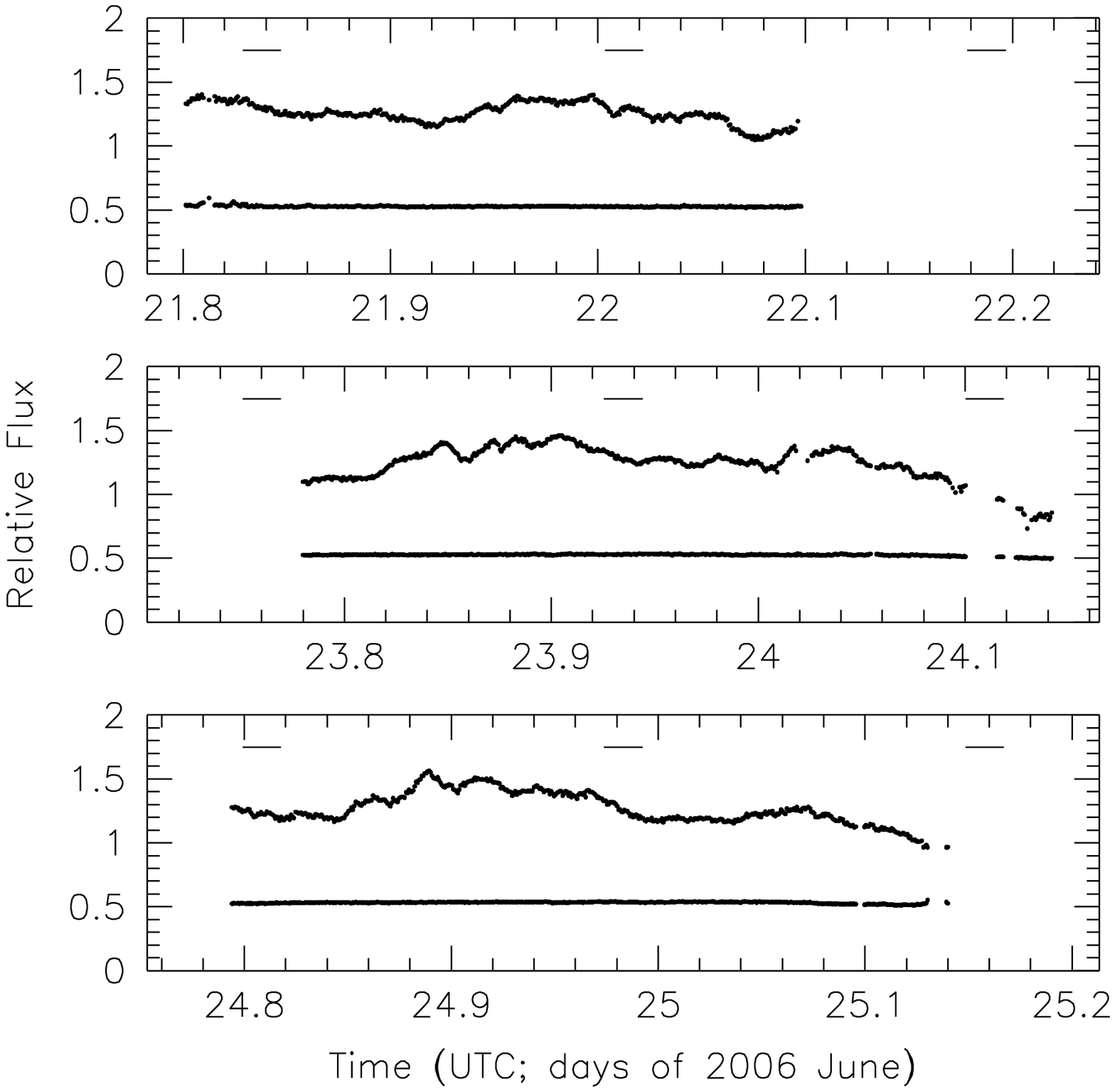}
\caption{Optical light curves of the optical counterpart of GX 9+9 and
a similar brightness check star.  The fluxes of the two stars are each
shown on an independent relative intensity scale; the scale for the
check star was chosen so that its light curves do not overlap with
those of the counterpart of GX 9+9.  These light curves were created
using only those measurements that deviated by less than 0.05 mag from
local running averages. One minute average intensities are shown.  The
observation times have been corrected to the Solar System barycenter.
The times of X-ray minimum intensity according to the ephemeris in
eq. (2) are indicated with horizontal bars.  For reference, 2006 June
20 is MJD 53906.
\label{fig:optlcs}}
\end{figure*}

For each image, we subtracted a bias frame and divided by a flat field
image prepared for each night. Since GX 9+9 is located between a
bright star and two close fainter stars, aperture photometry does not
yield reliable results in the presence of seeing variability. We
therefore used the DAOphot-II software package \citep{stet87,stet90}
to fit a point spread function (PSF) to each stellar image; the image
of the brightest star served as the PSF and photometric
reference. Another star of similar distance from the reference star
and similar brightness to GX 9+9 is used as a test of the photometric
systematics. It shows a typical uncertainty of 0.01\,mag (1\,$\sigma$)
during stable conditions. The photometric results were further
selected by rejecting those values which deviated by more than 0.05
mag from a 10 m local average and were then averaged in 1 m time bins.
The resulting light curves from the three nights with the best
conditions are shown in Figure~\ref{fig:optlcs}.

\begin{figure*}[tbh]
\centering
\epsscale{0.6}
\plotone{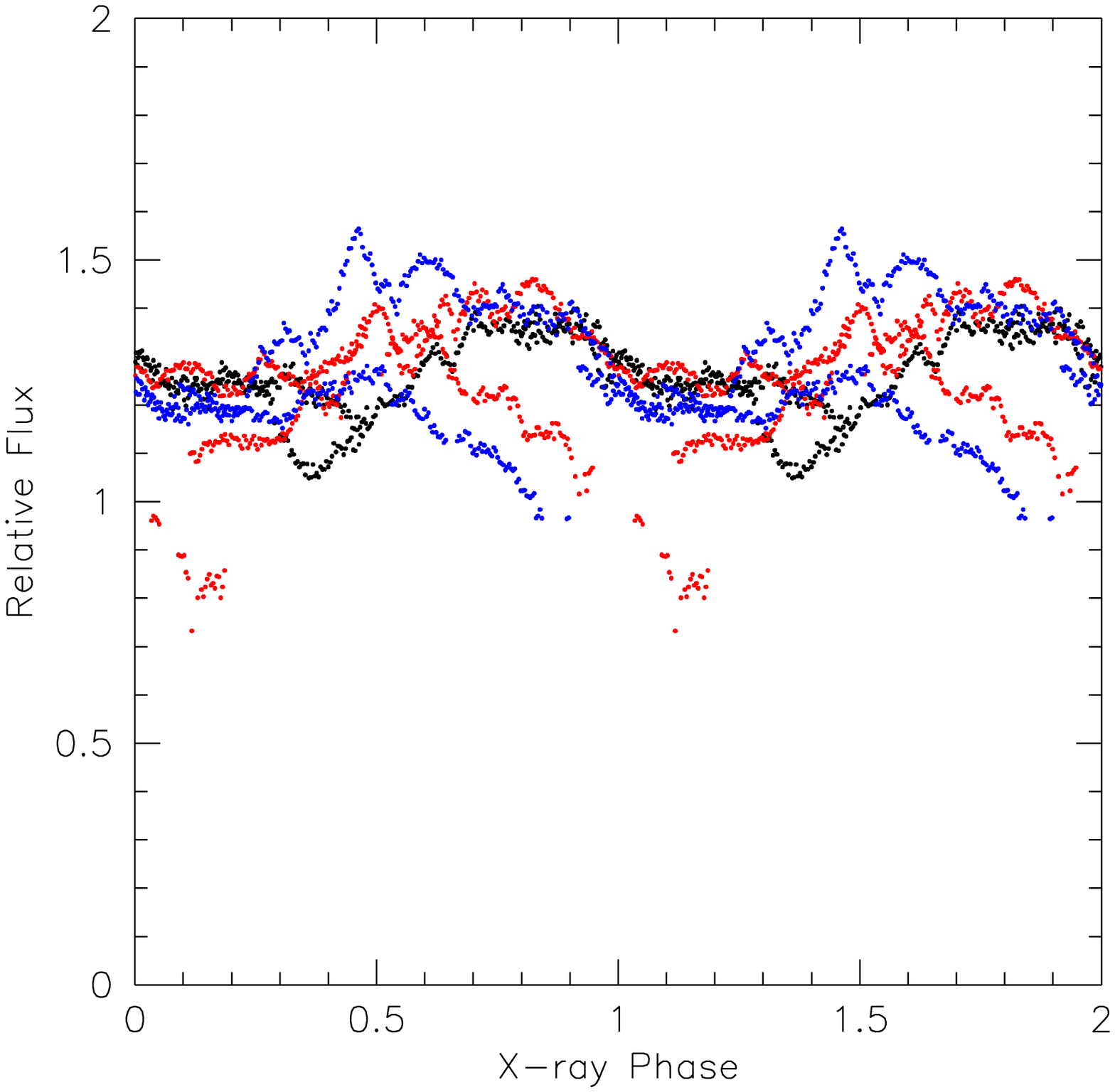}
\caption{The optical light curves of GX 9+9 from Figure 9 plotted
(twice) as a function of X-ray phase determined using eq. (2).  The
fluxes for each of the three nights (June 21-22, black; June 23-24,
red; June 24-25, blue) are shown on a normalized linear scale.  The
decline at late times of the intensity on the nights of June 23-24 and
24-25 may be an artifact (see text).
\label{fig:optfold}}
\end{figure*}

The light curves are plotted as a function of X-ray phase in
Fig.~\ref{fig:optfold}.  A marked decline in the intensity of the
counterpart is apparent late in the observations particularly on the
nights of June 23-24 and 24-25.  Thus the tails of the light curves
from these nights (shown in red and blue in Fig.~\ref{fig:optfold})
fall significantly below the other measurements at similar phases.
This could easily be an atmospheric effect since the counterpart of
GX~9+9 has a relatively blue spectrum compared to most stars deep in
the Galactic plane and, thus, may have more blue or UV flux than the
comparison star.  Furthermore, the field was on the meridian at
$\sim0.075$ days before $0^h$ UTC during these observations, and was
lowest in the sky near the end of the observations, especially on the
nights of June 23-24 and 24-25 (cf.  Fig.~\ref{fig:optlcs}).  This
also suggests that the use of the comparison star to normalize the
flux may not have removed other slow variations in the intensity of
the counterpart of GX~9+9. If we discount the tails of the red and
blue curves where they fall below the other photometric results, we
find that the variability has a strong periodic component.  The light
curves roughly suggest that optical minimum comes near X-ray phase
0.2, but the non-periodic aspects of the light curves do not allow a
firm conclusion to be drawn.

From a Lomb-Scargle periodogram of the data, we find the source varies
roughly sinusoidally with a period 4.17$\pm$0.11\,h, with shorter-term
variability superposed.  The amplitude of the 4.19 h modulation is
roughly $\pm15$\%.  This is close to or slightly smaller than that
seen in previous observations.

\section{Discussion} 

The long-term ASM light curve of GX~9+9 shows that its intensity has
changed slowly over the 12 years that it has been observed.  The form
of the variations suggest the presence of a periodicity with a period
of 1400 to 1600 days, but the light curve is not well-fit by a simple
linear function plus a sinusoid. Perhaps the variability is
quasiperiodic in nature.  Possible causes of the changes include 1)
variation in the accretion rate due to the donor star being a
long-period variable star that is undergoing small changes in size, 2)
the presence of a third star in the system in a $\sim 16$-day orbit
around the center of mass of the other two stars which produces the
variation by dynamical effects on the close pair
\citep[e.g.,][]{maz79,ford2000,zdz07}, and 3) the presence of a
tilted, precessing accretion disk (although the long period would be
hard to understand if this is the correct explanation).  In any case,
it would certainly be of interest to extend the X-ray light curve by
monitoring the intensity for many additional years. It would also be
of interest to determine whether similar variation is also manifest in
the optical band.

A detailed analysis of the ASM data has revealed significant changes
in the amplitude of the orbital (4.19 h) modulation in the X-ray
intensity.  When strong, the modulation is characterized by
energy-independent reductions in intensity that are limited in phase;
the light curves are definitely nonsinusoidal in form.

We used the ASM data to determine an ephemeris for the times of X-ray
minimum.  Our epoch differs from that of \citet{corn_etal07}; the
epoch in the latter report occurs at a phase of $\sim 0.24$ cycles
according to eq. (2).  In response to our query, we were informed
(R. Cornelisse, private communication) that the epoch of the time of
minimum that is explicitly stated in \S3.1 of \citet{corn_etal07} is
indeed incorrect but the X-ray phases used for the analyses and
figures in that paper actually had been calculated using the correct
epoch.

While the modulation is strong (MJD 53300-54100), the intensity of
GX~9+9 is reduced in comparison to that based on an interpolation from
the 2003-2004 time interval to the 2007 time frame.  However, the
degree of modulation during two similar reduced-intensity time
intervals (MJD 50300-51100 and MJD 51700-52500) was generally low
(with the exception of a small part of the latter interval when it was
of moderate strength as seen in interval 15 in Fig.~\ref{fig:lcpcs}).

PCA data obtained during the two-year interval when the X-ray
modulation was strong show evidence of dip-like intensity reductions
at orbital phases more or less consistent with the ASM folded light
curves.  The energy independence of the folded ASM light curves
(Fig.~\ref{fig:latefold}) is confirmed by the results in
Table~\ref{tbl-2} which show that the depths of the dip-like events
seen in the PCA data are roughly independent of photon energy.  The
intensity of GX~9+9 is generally less variable at low X-ray energies,
e.g., below 6 keV, than at higher energies.  This random variability
tends to mask the dip-like reductions at the higher energies.  Thus,
the reductions are more evident at low X-ray energies.  PCA and {\it
Chandra} data obtained prior to this two-year interval show that if
dip-like events were present, they tended to be rather shallow.  The
dip-like intensity reductions, regardless if they occurred during the
two year time interval when the modulation was strong on average or at
other times, are relatively limited in phase; they generally lasted
less than 0.1 orbital periods. On the whole, the PCA and {\it Chandra}
data suggest that the form of the light curves seen in the ASM data is
the result of the superposition of many dips with preferred but
varying phases, depths, and widths.

The amplitude of the 4.19 h modulation in the 2006 optical
observations is comparable to that seen in 1999 by \citet{kong06}, to
that seen in 2004 by \citet{corn_etal07}, and also to that seen in
1987 and 1988 by \citet{s90} even though the X-ray modulation strength
was much greater around the time of the 2006 observations than at the
time of the 1999 or 2004 observations.

The light curves from the 2006 optical observations, which covered
just over three orbital cycles, indicate that the time of minimum
optical flux is likely around 0.2 cycles after the time of minimum
X-ray flux given by eq. (2).  However, this conclusion about the
optical-to-x-ray phase relationship is not secure because of
deviations in the light curves from one cycle to another.  The light
curves obtained by \citet{kong06} from just over four orbital cycles
in 1999 present a similar picture; phase zero in their Figures 5 and 6
corresponds (by coincidence) to phase $0.01 \pm 0.20$ according to our
X-ray ephemeris. Note that all four nightly folded light curves in
their Fig. 6 have minima near phase 0.2. In contrast,
\citet{corn_etal07} obtained light curves over three orbital cycles in
2004 and concluded that the time of minimum optical flux (in the
continuum) corresponded closely to the time of minimum X-ray flux.
\citet{corn_etal07} did not show the individual light curves nor do we
know the precise time they used as the time of minimum X-ray flux.  It
is possible that the X-ray and optical light curves of GX~9+9 do not
maintain a strict phase relationship.  Extensive further observations
will be needed to resolve this question.

The X-ray intensity reductions in GX 9+9 are grossly similar to the
dips in 10 or more LMXBs such as 4U1915-05, X1254-690, EXO~0748-676,
X1755-33, or X1746-370 that appear at certain (limited) orbital phases
as absorption-like events.  In such dipping sources the dips do not
always occur at precisely the same orbital phases, often exhibit
complicated structure, and often are stronger at low energies in a
manner that suggests photoelectric absorption.  The dips occur with a
wide range of depths and can have durations up to as much as 25\% or
30\% of an orbital period.  At one extreme, the source 4U1915-05 has
exhibited dips that were essentially 100\% deep, showed evidence of
photoelectric absorption, and were, in some cases, temporally complex
\citep{walt82,whtswk82,chdot97}.  At the other extreme, the dips seen
in X1755-33 were not much more than 30\% deep, had essentially no
energy dependence, and, at least in some cases, were relatively simple
in their degree of structure \citep{whpar84,ch1993}. It is common for
the characteristics of the dips seen in a given source to vary; see,
e.g., \citet{bcs04} for a discussion of changes in the dipping
amplitude for X1746-370, \citet{parm86} and \citet{homan03} for
reports on dipping activity in EXO~0748-676, and \citet{sw99} for a
report on the entire cessation of dipping for a time in X1254-690.

No evidence of enhanced absorption at low energy is apparent in the
folded ASM light curves that show the GX~9+9 dips.  This is confirmed
by a few dips seen in PCA light curves which also show that the
intensity reductions do not have the degree of structure seen in
individual orbital cycles in many dippers.  However, the
characteristics of dips seen in various sources are rather diverse,
and the diplike events in GX 9+9 are reminiscent of the dips that were
evident in the {\it EXOSAT} observations of X1755-33
\citep{whpar84,ch1993}.

The dipping phenomenon has been explained at a general level as due to
the occultation of the X-ray source by localized regions of enhanced
density relatively far from the orbital plane around the place where
the stream of gas from the donor impacts the accretion disk
\citep{lubow76,whtswk82,walt82}.  According to this general idea, dips
will only be evident in those LMXBs whose orbital planes are inclined
to the line of sight in a relatively narrow range, namely those not at
such high inclinations that our line of sight to the neutron star
always intercepts the high density parts of the disk nor those at such
low inclinations that the line of sight is never intercepted by the
thick structures in the gas stream/disk collision region.  Since the
present results on GX 9+9 are similar to those previously seen in
X1755-33, they do not suggest extensions of the conventional picture,
but they do indicate that GX 9+9 is viewed at relatively high
inclination like the other dippers.  The energy independence of the
dips seen in GX9+9 may then be the product of different degrees of
energy-dependent absorption of multiple emission components such as
discussed by \citet{ch1993} in the context of X1755-33 rather than the
product of high ionization or elemental abundance anomalies
\citep{whpar84}.  However, one should note that there is not at this
time a full consensus on the interpretations of the observations of
dips especially on the locations and sizes of the emission components
and the degree of ionization of the absorbers \citep[e.g.,][and
references therein]{chbal04,chetal05,diaz06}.


\acknowledgments

We thank an anonymous referee for helpful comments.  ATS and MD
acknowledge support from the Spanish Ministry of Science and
Technology under the grant AYA\,2004--02646 and AYA\,2007--66887.

\end{document}